\documentclass[11pt,a4paper]{article} 
\pdfoutput=1
\usepackage[T1]{fontenc}

\usepackage{graphicx,amssymb,amsmath,latexsym} 
\usepackage{jcappubmy} 
\usepackage{color} 
\definecolor{nicered}{rgb}{0.7,0.1,0.1} 
\definecolor{nicegreen}{rgb}{0.1,0.5,0.1} 

\renewcommand\afterTocSpace{\vskip 4ex} 
\renewcommand\afterEmailSpace{\vskip25pt plus0.2fil minus12pt}
\renewcommand\afterTitleSpace{\vskip25pt plus0.2fil minus12pt}
\newcommand{\be}{\begin{equation}} 
\newcommand{\ee}{\end{equation}} 
\newcommand{\bea}{\begin{eqnarray}} 
\newcommand{\eea}{\end{eqnarray}}

\newcommand{\de}{\partial} 
\newcommand{\ha}{\frac{1}{2}} 
\newcommand\E{{\mathcal{E}}} 
\newcommand\K{{\mathcal{K}}} 
\newcommand\Q{{\cal Q}} 
 
\newcommand\T{{\cal T}} 
 
\renewcommand\P{{\cal P}} 
\newcommand\X{{\cal X}} 
\renewcommand\l{\lambda} 
 
\renewcommand\a{\alpha}

\newcommand\m{\mu} 
\newcommand\n{\nu} 
\newcommand\g{\gamma} 
 
\newcommand\s{\sigma} 
 
\newcommand\U{{\ensuremath{\cal U}}}

\renewcommand{\H}{{\mathcal H}} 
 
\renewcommand\l{\ensuremath{\lambda}}

\newcommand\ba{\begin{array}} 
\newcommand\ea{\end{array}}

\newcommand{\plm}{M_{\text{Pl}}^2} 
\newcommand\weff{w_{\text{eff}}}
 
 \title{Cosmology in General Massive Gravity Theories} 
 
\author[a]{D. Comelli,}
\author[b,c]{F. Nesti,}
\author[d,e]{L. Pilo} 
\affiliation[a]{INFN - Sezione di Ferrara,  I-35131 Ferrara, Italy}
\affiliation[b]{Gran Sasso Science Institute, viale Crispi 7, I-67100 L'Aquila, Italy}
\affiliation[c]{Ru\dj er Bo\v skovi\'c Institute, Bijeni\v cka cesta 54, 10000, Zagreb, Croatia}
\affiliation[d]{Dipartimento di Scienze Fisiche e Chimiche,
  Universit\`a di L'Aquila,  I-67010 L'Aquila, Italy}
\affiliation[e]{INFN, Laboratori Nazionali del Gran Sasso, I-67010
  Assergi, Italy} 
\emailAdd{comelli@fe.infn.it}
\emailAdd{fabrizio.nesti@aquila.infn.it}
\emailAdd{luigi.pilo@aquila.infn.it}

\abstract{
We study  the   cosmological FRW flat solutions generated in general massive gravity theories.
  Such a model are obtained adding to the Einstein General Relativity action a  peculiar non derivative potentials, function  of the metric components, that induce the propagation of  five  gravitational degrees of freedom. This large class of  theories includes both the case with a residual
Lorentz invariance 
as well as   the case with rotational invariance only.     It turns out that the Lorentz-breaking case is
  selected as the only possibility. 
  Moreover it turns out that that perturbations around strict Minkowski or dS space are strongly coupled. 
 The upshot is that even though  dark energy can be simply accounted by  massive
 gravity modifications, its equation of state $w_{\text{eff}}$ has to deviate from
  $-1$.  
  Indeed, there is  an explicit relation between the strong coupling
  scale of perturbations and the deviation of $w_{\text{eff}}$ from
  $-1$. Taking into account current limits on $w_{\text{eff}}$ and
  submillimiter tests of the Newton's law as a limit on the possible
  strong coupling scale, we find that it is still possible to have a
  weakly coupled theory in a quasi dS background.  Future experimental
  improvements on short distance tests of the Newton's law may be used
  to tighten the deviation of
  $w_{\text{eff}}$ form $-1$ in a weakly coupled massive gravity theory.}

\begin{document} 
 
\maketitle

 
\bigskip


\section{Introduction} 

Recently, a significative step forward in understanding massive gravity
was made in a series of papers~\cite{uscan,usweak,uslong} by the
nonperturbative construction of the most general theories with five
propagating degrees of freedom
(DoF).\footnote{See~\cite{Soloviev:2013mia} for a alternative analysis
  using Kuchar's Hamiltonian formalism.}  Besides its theoretical
interest, the main phenomenological goal is to investigate whether a
modification of gravity at large distances and a massive graviton can be
realized in a consistent theory which can also take care of the wealth of
other observational tests of gravity, from the smallest (submillimiter)
to largest (cosmological) scales.

In this work we study the cosmology of the massive gravity theories which
propagate five DoF in a systematic and model independent way. The
construction of~\cite{uscan,usweak,uslong} allows one to treat at once
theories which possess a residual Lorentz invariance as those with a
simpler rotational invariance.  Lorentz invariant massive
gravity~\cite{Gabadadze:2011} is phenomenologically not very successful:
in the ghost free version of massive gravity with graviton mass scale $m$
the energy cutoff $\Lambda_3=(m^2 \, M_{Pl})^{1/3}$ is too
low~\cite{HGS}, the theory is classically strongly coupled in the solar
system and, as already predicted in~\cite{HGS}, even the computation of
the static potential in the vicinity of the earth is problematic due to
possible large quantum corrections~\cite{pad}.  Cosmology is also
definitely troublesome: spatially flat homogenous
Friedmann-Robertson-Walker (FRW) solutions simply do not
exist~\cite{DAmico} and even allowing for open FRW solutions~\cite{open}
strong coupling~\cite{tasinato} and ghostlike  
instabilities~\cite{defelice} develop.  In the bigravity
formulation~\cite{DAM1,PRLus} FRW homogenous solutions do
exist~\cite{cosm}, however cosmological perturbations turn out to be
strongly coupled~\cite{cosmpert}.  On the other hand, things get better
if one gives up Lorentz invariance and requires only rotational
invariance~\cite{Rubakov,dub,usweak}. Within the general class of
theories which propagate five DOF found in~\cite{uscan,usweak,uslong}, in
the Lorentz breaking (LB) case most of the theories have much safer
cutoff $\Lambda_2 =(m \, M_{Pl})^{1/2}\gg \Lambda_3$ and also avoid all
of the phenomenological difficulties mentioned above. A recent
comprensive  review of massive gravity can be found in~\cite{deRham-rev}. 

We study, in full generality, the conditions for the existence of a
homogeneous flat FRW cosmological background.
The massive deformation of general relativity show up as an additional effective energy momentum tensor
whose conservation has a crucial impact on the behaviour of perturbations.  In particular, the present accelerated de Sitter
(dS) phase of the Universe can be naturally accounted for by massive gravity, but it turns out that
the effective equation of state of dark energy is directly connected to both the large distance
scale of modification of gravity and the strong coupling scale of gravitational perturbations around
quasi dS space.

The outline of the paper is the following. In section \ref{genmgr} we briefly recall the
construction of the general massive gravity theories with five DoF. The existence of background FRW
solutions are considered in section \ref{backcosm}, where the conditions due to the Bianchi identity
are studied and where we discuss the effective perfect fluid resulting from the massive deformation
of gravity; some observational constraints are also discussed. Perturbations around FRW background
and the relation with the effective gravitational dark fluid equation of state are discussed in
section \ref{perts}. Section \ref{conc} contains our conclusions.

\section{Massive gravity with 5 DoF}
\label{genmgr}
Generic nonderivative deformations of GR are defined by adding to the Einstein-Hilbert action a
potential $V$ which depends on the metric $g_{\m\n}$,
\be
S= \plm \int d^3x \sqrt{g}\,\Big[ R - \, m^2 \, V (g) \Big] \, ;
\label{eq:act} 
\ee 
the parameter $m$ sets the scale for the graviton mass.  The general features of such gravity
modification were studied~\cite{uscan,usweak,uslong} at the nonperturbative level by using
Hamiltonian analysis. The ADM decomposition~\cite{ADM} of the metric reads
\be \label{adm}
g_{\mu \nu} = \begin{pmatrix} - N^2 + N^i N^j \g_{ij} & \g_{ij}N^j \\ 
\g_{ij}N^j  & \g_{ij} \end{pmatrix} \, ,
\ee 
and the potential $V$ can be regarded as a function of lapse $N$, shifts $N^i$ and spatial metric
$\g_{ij}$.\footnote{The action (\ref{eq:act}) is written in the so called unitary gauge, which we
  use throughout our work.  By using a set of four additional (St\"uckelberg) scalar fields, $V$ may
  be written as diffeomorphism invariant scalar function, see the detailed discussion
  in~\cite{uslong}, and general frameworks in~\cite{HGS,dub,Rubakov:2008nh}. By definition in the
  unitary gauge the derivatives of the St\"uckelberg fields are trivial; as a result, $V$ is
  function of the ADM variables only.}
 
In~\cite{usweak,uslong} the most general potential that propagates five DoFs at nonperturbative
level was found, under the requirement that rotations are unbroken.  It turns out that $V$ is
parametrized in terms of two arbitrary functions\footnote{Note, while in~\cite{usweak} the potential
  was parametrized in terms of \emph{three} functions, in the following paper~\cite{uslong} it was
  shown that by solving the associated Monge-Ampere problem one of them is just a constant of
  integration, so that \emph{two} functions only are sufficient. See also footnote 6 on page 10
  in~\cite{uslong}.}  $\U$ and $\E$ of $\g^{ij}$ and a set of new shift variables $\xi^i$ which are
related to $N^i$ through the implicit relation
\be 
N^i - N \;\xi^i= \;\U^{ij}\; \E_{\xi^j}\equiv \Q^i(\g^{ij},\xi^i)\,.
\label{eq:xi} 
\ee 
Here, $\E$ is a generic function $\E(\g^{ij},\xi^i)$ of the spatial metric $\g^{ij}$ and $\xi^i$,
while $\U$ is an arbitrary function of the special combination of variables $\g^{ij}-\xi^i\,\xi^j$,
namely
\be\label{kk}
\U(\K^{ij}),\qquad 
\K^{ij}=\g^{ij}-\xi^i\,\xi^j  \,.
\ee 
The notation $\E_{\xi^i}$ denotes the partial derivative with respect to $\xi^i$, and $\U^{ij}$ is
the inverse of the Hessian matrix $\U_{ij}=\U_{\xi^i\xi^j}$.  The bottom line of the
canonical analysis if that all potentials which propagate five DoF are of the form
\be
 V(N,N^i,\g_{ij})=\U+N^{-1} \left(\E+ \Q^i\; \U_{\xi^i} \right),
\label{pot} 
\ee 
which we will use in this work.  

\pagebreak[3]

The general construction of $V$ is rather powerful: a whole set of interesting physical implication
can be worked out without even specifying the form of $\U$ and $\E$. It is worth to stress that the
Lorentz invariant ghost free massive gravity theory found in~\cite{Gabadadze:2011} is of course of
the form (\ref{pot}), see~\cite{uslong} for the details.\footnote{One can further take the extra
  metric dynamical by introducing an additional Einstein-Hilbert action, see for
  instance~\cite{DAM1, PRLus, myproc, spher, energy, spher1}, leading to a bimetric theory. We will
  not discuss this option here.}  As studied in~\cite{usweak,uslong}, besides the Lorentz-invariant
case, a whole new class of interesting theories which are weakly coupled at phenomenologically
interesting scales can be constructed.
The existence of ghost-free weakly coupled massive gravity theories with only rotational invariance
can also be important to avoid the issue of acausal propagation~\cite{Deser:2012qx} which affects
the already troubled Lorentz-invariant case, and appears to be due to the strong nonlinearities
present there.

\section{FRW Background and the Bianchi identity} 
\label{backcosm}

All observations~\cite{hogg} are consistent with the cosmological principle that
we can slice our spacetime in spatial homogeneous and isotropic hypersurfaces
associated with observers (cosmological observers) for whom the CMB is
practically isotropic.  Generically, massive modified gravity contains {\it
  apriori} nondynamical metric(s), much like the Nordstrom theory of gravity.
The construction recalled above, allowing for Lorentz-breaking scenarios, relies
mainly on a fiducial 3d metric $\delta_{ij}$ (see section~6
in~\cite{uslong}). As a result, there is also another special class of observers
(preferred frame observers) for which some the dynamical metric has a preferred
form.  Whether the cosmological and fiducial frames coincide is a physical
hypothesis.
Working in a flat homogeneous space\footnote{We consider a flat
  spatial section, taking into account the overwhelming observational
  evidences in its favor. We leave the generalization to the non flat
  case for a future work.  As well know there are no cosmological
  solutions for a flat FRW universe \cite{DAmico}, while there are
  solutions for the open ($k=-1$) FRW cosmology~\cite{open}.
  Incidentally, we recall that for this Lorentz invariant cases, even
  resorting to nontrivial St\"uckelberg fields one has to face strong
  coupling and ghost instabilities~\cite{tasinato,defelice}, for a
  review see~\cite{defelice-rev}.  }  (with cartesian coordinates for
simplicity) it is possible to bring, via a time redefinition, the
4-dimensional metric in the form
\be\label{aa} ds^2=-N(t)^2\;dt^2+a(t)^2\;(d\vec
x)^2\,,
\ee 
that matches with the ADM form of eq.~(\ref{adm}). The lapse can be interpreted
as the presence of a nontrivial temporal Stuckelberg field, making explicit time
reparametrizations. On this background, we have $N^i=\xi^i=\Q^i=0$, while in
general one can have $\partial_{N^j}\xi^i\neq 0$ and $\partial_{\xi^j}\Q^i\neq
0$.  We also have $\U_{\g^{ij}}\propto \g_{ij}$ and we define the scalar
quantity $\U'$ by $\U_{\g^{ij}}\equiv \U'\;\g_{ij}$, and similarly for $\E'$ by
$\E_{\gamma^{ij}}= \E' \; \gamma_{ij}$.  One also has
$\U_{\xi^i\xi^j}=-2\;\U'\;\g_{ij}$ and finally we use ${H}=\dot a/a$.
 
So our main hypothesis is that we preserve homogeneity and spatial isotropy both
at the level of physical metric that for the non dynamical metric 
Notice that also that homogeneity and isotropy forces the flat massive gravity
preferred frame and our cosmological CMB frame to coincide.

The Einstein equations take the form
\be
E_{\mu \nu} =  \T_{\mu \nu} + 8 \pi G \, T^{(\text{matt})}_{\mu \nu}\, ,
\ee
where $T^{(\text{matt})}_{\mu \nu}$ is the matter energy momentum tensor (EMT) and $\T_{\mu \nu}$ is the
contribution of the potential $V$. In particular, we have
\bea 
\T_{00}&=& m^2\,\frac{N^2}{2}\;\U\equiv \rho_{\text{eff}} \, N^2\;\nonumber\\
\T_{ij}&=&m^2 \,\g_{ij}\, \left[ \U'- \frac{\U}{2}+\frac{1}{N}\left(\E'-\frac{\E}{2}\right)
\right] \equiv  p_{\text{eff}} \, \gamma_{ij} \, ,
\label{eq:backgrEE}
\eea
where we used the fact that $\partial_t \U = \U_{\g^{ij}}\partial_t\g^{ij} =
\U'\;\g_{ij}\partial_t\g^{ij}=- 6 \,H\, \U' $.

We assume that matter is minimally coupled to gravity respecting general covariance, so that
$T_{\m\n}^{(\text{matt})}$ is separately covariantly conserved. Therefore, also $\T_{\mu \nu}$ has
to be separately conserved. The resulting Bianchi identity, $\nabla^{\nu}\T_{\m\n}=0$, takes the
simple form
\bea
 \de_t \rho_{\text{eff}} + 3 H (\rho_{\text{eff}}
+p_{\text{eff}}) =0
\quad \to  \quad 
 H\left.\left(\E'-\frac{\E}{2}\; \right)\right|_{\xi^i=0}=0 \, .
\label{bianchi}
\eea
Physically this condition can also be understood as the implementation of time
reparametrization. While the $\U$ part is automatically conserved (accordingly
to time reparametrization $\U$ appears linearly in $N$ as $N\U$ in the action),
the $\E$ part is constrained by this equation. In fact, (\ref{bianchi}) has far
reaching consequences. It implies that either $H=0$ (i.e.\ $a$ is constant in
time) or that the second factor vanishes automatically, independently of $a$
(otherwise it would imply an algebraic constraint on $a$, clearly incompatible
with any sensible cosmology). Its automatic fullfillment poses a strong condition
on the function $\E$, on the surface $\xi^i=0$, which has to be satisfied by
choosing some particular form of $\E$.  For instance, one can take $\E$ to be
homogeneous of degree $-3/2$ in $\g_{ij}$,
\be
\E =\g^{-1/2}\;{\P}\left(\frac{\text{Tr}[\g^2]}{\text{Tr}[\g]^2},\frac{\text{Tr}[\g^3]}{\text{Tr}[\g]^3},\xi^i\right)
\ee
with any function ${\P}$, but other choices are possible.

Incidentally, we remark that the form of $\E$ dictated by the dRGT
Lorentz-invariant massive gravity, $\E_{(LI)}=(1-\xi\g\xi)^{-1/2}$
(see~\cite{uslong}) does not satisfy~(\ref{bianchi}) and thus no sensible
cosmology is possible, as it was already realized in~\cite{DAmico}.
\footnote{Such a term corresponds to the potential $(Tr[X^{1/2}]-3)$, one of the four possible  pieces of the dRGT potential , see \cite{Gabadadze:2011} and \cite{uslong}.}
 We are thus
led to the conclusion that Lorentz-breaking is necessary, in massive gravity
with 5 DoF, to admit a nontrivial cosmology for a spatially flat
Universe. 

Before moving on, it is worth mentioning that our framework is the
most general which preserves spatial homogeneity and isotropy, and has
SO(3) invariance in the theory. In fact, at most one might promote the
3D fiducial metric $\delta_{ij}$ used in the lagrangian to a
time-dependent one, $\zeta(\tau) \delta_{ij}$, with $\zeta$
arbitrary. However, in this case the Bianchi identity becomes
\be 
{\cal
  H}\left(\E'-\frac{\E}{2}\right)+\frac{\dot\zeta}{\zeta}\;\U'=0\,.
\ee 
So, instead of constraining the potential, the Bianchi identity results (for
$\U'\neq 0$) in an algebraic equation that fixes the time expansion $a(\tau)$ of
our universe in terms of the absolute field $\zeta(\tau)$.  We thik this scenario
to be unphysical, in that we feel uneasy in relating time a dependent function
introduced as a supposedly nondynamical fiducial metric to the physical scale
factor that should depend on the matter content. For such a reason we do not
consider this possibility further.

\paragraph{Generic Background evolution.}
In the following we will assume that the Bianchi condition is automatically satisfied, with
a nontrivial $H$, thanks to some particular choice of $\E$.  In this case, $\E$ effectively drops
out of the background equations~(\ref{eq:backgrEE}), and the contribution of $V$ to the Einstein
equations has the form of a ``gravitational'' perfect fluid, with effective density and pressure
given by
\be
\label{fluid}
\rho_{\text{eff}} \equiv m^2 \, \U \, , \quad
 p_{\text{eff}} \equiv   \, m^2 \left(2\, \U' - \U  \right) .
\ee
Thus, the effective  gravitational fluid mimics the following equation of state
\be
\label{eq:weff}
\weff=\frac{p_{\text{eff}}}{\rho_{\text{eff}}}=-1+\frac{2\,\U'}{\U}\, .
\ee
When $2\U'/\U <1$, the gravitational fluid mimics dark energy; in particular if $2\U'/\U <0$, one
has $w<-1$, turning the gravitational fluid into a phantom one.

The fluid contributes to the expansion rate through the standard equation
\be
3 H^2=N^2 \left(\frac{m^2 \, \U}{2} + 8 \pi G \,  \rho_m \right) \,.
\label{eq:H}
\ee
Now, given that the function $\U$ is still generic, it can accomodate the most diverse
cosmologies. In fact, it is sufficient to observe that given \emph{any} cosmological history in
terms of a positive $H(t)$ or equivalently in terms of $H(a)$, together with the matter content in
terms of $\rho_m(a)$, the previous equation can always be solved by choosing a particular function
$\U(\K)$ so that on the background $\K^{ij}=a^{-2}\delta^{ij}$ it gives the correct $a$-dependence
for (\ref{eq:H}).  Clearly, any bizarre accelerating or decelerating or oscillating and bouncing
cosmology can be present.


\paragraph{Constraints from the present epoch and early time.}
Having clarified that any cosmological evolution can result from the choice of the function $\U$,
one can analyze the constraints on $\U$ from present day ($a\sim1$) and early time ($a\ll1$)
observational constraints, under some mild hypotheses for its form.

For instance, we can consider the early or late time expansions as $\U=\sum_{n=0}^{\infty}
\tilde\U_n\,a^{-n}= \sum_{n=0}^{\infty} \bar\U_n\,(a-1)^n$, and assume there are no strong
cancelations between the coefficients in the series. This allows us to put constraints on the
coefficients $\tilde\U_n$ and $\bar\U_n$.

If the effective gravitational density is driving the present day observed acceleration, the
background energy scale of the potential is fixed as (here $8 \pi G=1/2\plm$)
\be
V= m^2M_{Pl}^2\,\bar\U_0\sim \Omega_{\Lambda} \rho_c  
\qquad \to \qquad
 m^2\, \bar\U_0\sim H_0^2 \sim \left(10^{-34} \, {\rm eV}\right)^2
\label{scale}
\ee
with $H_0$ the present Hubble parameter. The effective equation of state can also be expanded, as
\bea\nonumber
\weff&=&-1-\frac{1}{3}\;\frac{\bar
  \U_1}{\bar\U_0}-(1-a)\frac{\bar\U_1^{2}-\bar\U_0\;(2\;\bar\U_2+\bar\U_1)}{3\;\bar\U_0^2}+\cdots\\  
&\equiv& w_0+(1-a)w_a+\cdots\,,  
\eea
where $w_0$ is the present value of $\weff$ while $w_a$ represents a possible time evolution of
the equation of state.  The combined data of Planck, WMAP low-multipole (WP) and baryonic acoustic
oscillations (BAO)~\cite{planck} give the following conservative observational constraints (at
$95\%$ C.L.)
\bea
&&w_0 = -1.04^{+0.7 }_{-0.7}\,,\qquad w_a<1.32\,, \nonumber 
\\[1ex]
{\rm i.e.}\quad &&\frac{\bar\U_1}{\bar\U_0} = \,0.12\pm2.1  \,,\qquad
 \frac{\bar\U_2}{\bar\U_0}<\, 2\pm3\,.
\eea
On the other hand, the global analysis including supernovae data (fig.\ 36 in~\cite{planck}) has a
mild preference for $w_a\simeq-1.6$, and $0$ is disfavoured at $2\sigma$. Any future hint like 
redshift depedent  equation of state for dark energy can find in the massive gravity cosmology a simple explanation in
terms of the nontrivial form of $\U$. 

Focusing instead on the early Universe, each coefficient $\tilde \U_n$ mimics a different fluid with
an equation of state $w_n=-1+n/3$.  For instance, $\tilde\U_1$ gives dark energy with $w=-2/3$,
$\tilde\U_2$ behaves like spatial curvature , $\tilde\U_3$ like non relativistic matter and
$\tilde\U_4$ like radiation.  Defining the dimensionless constant $x\equiv
m^2M^2_{Pl}/\rho_c=\Omega_\Lambda/\bar\U_0$, if we impose that the gravitational fluid does not
alter the background evolution from big bang nucleosynthesis (BBN) on, a quick estimate assuming no
cancelations between coefficients gives the following bounds
\be\label{constr}
\begin{split}
& x\; \tilde\U_{2}  \leq \Omega_{K}\sim10^{-2},\\
& x\;  \tilde\U_{3} \leq
\Omega_{\text{mat}}\sim 10^{-2} \,,\\
& x\; \tilde\U_{4} \leq \Omega_{\text{rad}} \sim 10^{-5}\,,\\
& x\; \tilde\U_{4+n} \ll z_{\text{bbn}}^{-n} \, \Omega_{rad} \sim 10^{-(5+8n)}\,, \qquad   (n\ge0)\,,
\end{split}
\ee
where $z_{\text{bbn}} \sim 10^8$ is the redshift at BBN. Note that higher
coefficients are much more constrained because their scale factor dependence would make their
contribution more dominant at early times.

\medskip

In the next section we discuss the minimal conditions under which the perturbations around the
cosmological background and flat space are well behaved.

\section{Impact on behaviour of perturbations}  
\label{perts}

Exploiting the construction of the massive gravity potential we can study some very general aspects
of perturbations around FRW and Minkowski backgrounds.  Let us start by expanding the metric around
the FRW background (\ref{aa}), choosing $\bar N=a$ (conformal time), so that the
background metric is conformally flat:
\be
g_{\mu\nu}=\bar
g_{\mu\nu}+a^2 \, h_{\mu\nu}=a^2 \left(\eta_{\mu \nu} + h_{\mu \nu}
\right) \, .
\ee
The derivatives with respect to conformal time is denoted by $'$, in particular the conformal Hubble
parameter will be denoted by ${\cal H}\equiv a'/a$ and the standard one is $H\equiv a'/a^2$. It is
also convenient to decompose the perturbations according their transformation properties under
rotations
\be
\begin{split}
&h_{00} = \psi \, , \qquad h_{ij}= \chi_{ij} + \de_i s_j + \de_j s_i + \delta_{ij} \, \tau +
\de_i \de_j \sigma , \qquad h_{0i} = u_i +
\de_i v  \,,
\label{dec}
\end{split}
\ee
with $\chi_{ij} \delta^{ij} = \de_j  \chi_{ij} = \de_i s_i = \de_i u_i =0 $.  

Consider the case of a Universe dominated by dark energy induced by the massive gravity potential
$V$. The quadratic action for the perturbation can be computed for a generic $V$ and details are
given in appendix~\ref{appquad}.  Particularly useful is the quadratic expansion of the deforming
potential parametrized in terms of the masses $m_{0}$,\ldots $m_4\sim O(m)$ defined by~\cite{diego}
\be
\plm \, \sqrt{g} \, (m^2 \, V- 6 H^2) \equiv 
\frac{a^4M_{Pl}^2}{4} \Big(m_0^2\, h_{00}^2+2\,m_1^2\, h_{0i}^2   
    -2\,m_4^2 \,h_{00}h_{ii} +m_3^2 
\,  h_{ii}^2  - m_2^2\, h_{ij}^2\Big) \,.
\label{masspar}
\ee
In particular, for a general massive gravity modification of GR,  one has the following result
\be
m_0^2=0 \,  ,\qquad 
m_1^2= 2 \, m^2 \, \U' \X\, , \qquad m_4^2 = m^2 \, \U'
\, , \qquad \text{with } \X=  \frac{2 a \U'}{2 a \U'  + \E_v''} \,,
\label{masses}
\ee
where we defined $\frac{\de^2 \E}{\de \xi^i \de \xi^j}_{| \bar g}= \E''_v \bar \g_{ij} $.  The
expressions for $m_2$ and $m_3$ are not needed here and are reported in
appendix~\ref{appquad}. Using (\ref{masspar}), the complete analysis of perturbations around a FRW
background in massive gravity was given in~\cite{diego}. Around a FRW background ~\cite{diego}, as
well as around flat space~\cite{Rubakov,dub,PRLus}, when $m_0=0$ and $m_1 \neq 0$ the number of
propagating degrees of freedom at linearized level is five; namely two tensors $(\chi_{ij})$, two
vectors $(s_i)$ and a scalar $(\sigma)$.  Thus, the above result $m_0=0$ is in agreement with the
nonperturbative canonical analysis, and the linearized approximation captures all the propagating
DoF.  This is a sign that the system can be analyzed by weak-field expansion (see also below).

\medskip

The conditions for UV stability (no ghosts) are directly dictated by $m_1^2$ and
$m_4^2$~\cite{diego}, and turn out to be the following
\be
\E_v'' > -2a\U'\,,\qquad \E_v''(\E_v''-2a\U'+2a) >0\,,
\ee
which can be always satisfied by choosing $\E$ such that $\E''_v$ is sufficiently large and
positive.  Thus we conclude that the generic massive gravity theories are free of ghosts with mild
assumptions on $\E$ and $\U$.

\medskip

\paragraph{Strong coupling.} It is well known that the violation of diffeomorphisms by the potential
implies that the new propagating modes can become strongly coupled at some energy or momentum scale.
This can manifest already at the classical level for instance when perturbation series break down,
and also at the quantum level, where in the spirit of the effective field theory, the possible
effective operators suppressed by a cutoff scale become important.  Let us first discuss the
classical case.

In GR the classical perturbation expansion breaks down in the presence of a source of mass $M$ at
the Schwarzschild radius $r_s\sim M/M_{Pl}^2$. In massive gravity, since new fields mediate the
gravitational interaction, one can expect deviations from the above behaviour. For instance, in
Lorentz invariant (LI) massive gravity~\cite{Gabadadze:2011}, the perturbation expansion breaks down
at a distance (Vainshtein scale) $\Lambda_* = (M_{Pl}m^2)^{1/3} (M/M_{Pl})^{1/3}$ from a source of
mass M, which becomes untenable already for macroscopic objects like the Earth or Sun.

In the Lorentz breaking case the situation is completely different, and there is in fact no
classical strong coupling apart from the Schwarzschild radius as in
GR.  To see this, we note that, for static
configurations and also at  large spatial momenta,  the auxiliary
fields  vanish~\cite{diego} and one can study interactions in terms of the sole physical propagating fields
$\chi_{ij}$, $s_i$, $\sigma$. Their quadratic Lagrangian in the UV limit is given by
\bea
\label{eq:Lsigma}
L_\sigma &=& a^4 \, \Lambda_2^4 \, \nabla^2 \left(\frac{\U'}{2} \, \sigma'^2-\frac{\l_{23}^2}{2} \nabla^2 \sigma^2 \right),\qquad \l_{23}^2\equiv\frac{m_2^2-m_3^2}{m^2}\,,\\
\label{eq:Ls}
L_{s_i} &=& a^4 \, \Lambda_2^4  \left(\frac{2\U'\X}{2} \, {s_i'}^2-\frac{\l_{2}^2}{2} \nabla^2 s_i^2 \right),\quad\qquad \l_{2}^2\equiv\frac{m_2^2}{m^2}\,.
\eea
where $\Lambda_2^2=mM_{Pl}$ and we are using the spatial Fourier transform, defining
$\nabla=\sqrt{-\Delta}$. Notice also  that for for $\E=Q^i_{\xi^j}=0$, one has $\X=1$.

Suppressing for a moment dimensionless factors, these fields are made canonical as follows
\be
\sigma^c = \Lambda_2^2 \nabla \sigma \,,\qquad 
s_i^c = \Lambda_2^2  s^i \,,\qquad 
\chi_{ij}^c = M_{Pl}\, \chi_{ij} \,.
\ee
We can then study a generic Lagrangian term
\be
\label{eq:terms}
\Lambda_2^4\,(\nabla^2 \sigma)^n\,(\nabla \,s)^r\,\chi^p\sim
 \frac{(\nabla \sigma^c)^n \,(\nabla \,s^c)^r\;(\chi^c)^p}{\Lambda_2^{2(n+r)-4}\,M_{Pl}^p} 
\ee
and we see that the leading operators are the ones with $p=0$.  Focusing on $\s^c$ (the same holds
for $s_i^c$), around a classical static source, the perturbative expansion works (no strong
coupling) as long as
\be
 \nabla\sigma^c< \Lambda_2^2=m\,M_{Pl}\,.
\ee
However, a crucial fact is that the canonical field $\sigma^c$ induced by a static source
$T_{matter}=M\,\delta^{3}(r)$ is given by, after integrating out the auxiliary fields,
\be
\nabla^2\sigma^c-\;\frac{m}{M_{Pl} \nabla}\, T_{matter}=0\quad \to\quad  \nabla\sigma^c\sim \frac{m\,M }{ M_{Pl}\, r}\,.
\ee
As a result, $\Lambda_2$ cancels out and one can use the weak-field expansion outside a static
source at distances larger than a critical radius $r_c$ which is simply given by
\be
 r_c \simeq r_s=\frac{M  }{M_{Pl}^2}
\ee
i.e.\ the same as in GR. We conclude that the theory does not suffer from new classical
nonlinearities near a source, and the perturbation series just break down at the Schwarzschild
radius, as in GR. Notice that, at this radius all Lagrangian terms (\ref{eq:terms})\ become
important.  If one were to ignore quantum effects (see below), the perturbation series would be
reliable even at distance scales smaller than $1/\Lambda_2$ (provided this were still larger than
the Schwarzschild radius). This happens because the fields responsible for the strong coupling,
$\sigma$ and $s_i$, are not directly sourced by matter but only via a coupling with the standard GR
fields, and the coupling itself is $m$-suppressed.

The absence of a Vainshtein strong coupling scale and the disappearance of $\Lambda_2$ from the
classical perturbation series was also explicitly demonstrated with the first perturbative orders
in~\cite{usweak}.  This result is a remarkable fact, because it suggests that as a classical theory
the present theory can be used perturbatively also at short distances.

Keeping track of the dimensionless coefficients in~(\ref{eq:Lsigma}) and (\ref{eq:Ls}) does not
alter the result, even in the limit of small $\U'$ as required by cosmology.  In fact, for static
configurations the time-derivative kinetic term will clearly be less important.  A possible
worsening of the nonlinearity scale, i.e.\ a larger Schwarzschild radius, might be present if
$\l_{23}$ is taken to be parametrically~small.

\medskip

A consequence of the requirement of small $\U'$ is only present for time dependent classical
solutions. In fact, the vanishing of the temporal kinetic terms for $\s$ and $s_i$ for $\U'\to0$
(see (\ref{eq:Lsigma}) and (\ref{eq:Ls})) tells us that, because the canonical analysis still
requires five DoF, in the limit of $\U'\to0$ these modes must propagate through interaction terms,
i.e.\ become progressively strongly coupled.  This will reflect on the classical response to highly
oscillatory source, with frequency $\omega$ much larger than momentum $k$ (and of course than $H$,
$m$).

\medskip

Let us now consider massive gravity as a quantum theory.   In the spirit of effective field theory, a
cutoff $\Lambda$ is expected.  $\Lambda$ is the scale where we loose control of the theory and some
UV completion becomes mandatory to make any physical prediction.  In GR such scale is the the Planck
mass, with no foreseeable phenomenological consequences.  In the generic completions of the
Fierz-Pauli massive gravity, the cutoff is as small as $\Lambda_5 = (M_{Pl}m^4)^{1/5}$~\cite{HGS},
or at best as in the dRGT theory $\Lambda_3 = (M_{Pl}m^2)^{1/3}$, at the price of an infinite number
of fine tunings~\cite{HGS}.  If $m \sim H_0$ then $\Lambda_3 \sim (1000 \text{Km})^{-1}$, so that
one looses control at macroscopic distances and even the static gravitational force in experiments
around the earth is incalculable~\cite{pad}.

For the class of potentials (\ref{pot}) analyzed here with rotational invariance only, the situation
is much more favorable with a reasonable cutoff $\Lambda_2= (M_{Pl}m)^{1/2}\gtrsim
(10^{-4}\,\text{mm})^{-1}$ in flat space~\cite{usweak}.  This can be seen directly from the
canonical fields or the Lagrangians (\ref{eq:Lsigma}) and (\ref{eq:Ls}).  From those expression is
also clear that taking a progressively small $\U'$ will worsen the quantum cutoff.  Indeed, without
doing any actual loop computation, by rescaling time by a factor $\sqrt{\U'}$, the troublesome small
dimensionless parameter is removed from the Lagrangian and the neat effect is to replace
$\hbar\to\hbar/\sqrt{\U'}$ in the exponential of the action. Similar result can be understood by
rescaling energy in the loops.  As a result, the loop expansion will become less convergent.

Using standard arguments~\cite{Georgi:1985kw} one knows that given a (renormalized) classical
nonlinearity scale $f$, the quantum corrections tend to become important at the scale $4\pi f /
\sqrt{\hbar}$ (provided the generated operators contain even powers of a cutoff, as
in~(\ref{eq:terms})). In our case the scale of classical nonlinearity is $\Lambda_2$, and together
with the effect of $\U'$ on $\hbar$, we expect a quantum cutoff of the order of $\Lambda\simeq 4\pi
\Lambda_2 (U')^{1/4}$.  To see quantitatively the effect, we can use the effective equation of state
for the gravitational fluid (\ref{eq:weff}) to solve for $\U'$ and use the background equation
$m^2\U=H^2$ to recast the cutoff $\Lambda$ in terms of the deviation of $\weff$ from~$-1$:
\bea 
\Lambda \simeq 4\pi \sqrt{M_{Pl}H}\, (\weff+1)^{1/4} \,.\qquad 
\label{cutw}
\eea
Thus, the strong coupling scale is directly linked to $\weff+1$, and note that the graviton mass
disappears in favour of the explicit appearance of $\weff$.  In any given massive gravity theory
with five DoF, an accelerated expansion phase can exist, but as soon as the equation of state gets
close to the DeSitter phase with $\weff \approx -1$, gravitational perturbations tend to become
progressively strongly coupled.\footnote{Clearly, for particular models the actual loop expansion
  could be still more convergent by virtue of the particular operator coefficients. The
  model-dependent analysis goes beyond the scope of this work.}

How safe we are from large quantum effects depends on how far $\weff$ is from $-1$ and at the same
time on the value of $m$.  Note that $4\pi\sqrt{M_{Pl}H}$ at present time gives a cutoff at
distances of the order of $10^{-2}\,$mm at which no deviations from the Newton law have been
found~\cite{newton}.  The present uncertainty on the deviation of $\weff$ from $-1$ is still of
order one~\cite{planck}, and one can still be consistent both with experiments at small scales (test
of the Newton's law) and at large scales (cosmology).  However, (\ref{cutw}) gives an intriguing
connection between small and large scales, and future progress on the determination of $\weff$ and
on short distances test of gravity will turn (\ref{cutw}) into a prediction, if one want to keep the
theory calculable.

\medskip

Finally, a brief comment comparing with previous analyses of massive gravity is in order. As we
mentioned in the introduction, in all earlier approaches, homogeneous cosmology was always
problematic: in the Lorentz invariant case it simply does not exist~\cite{DAmico}, while in the
bigravity approach~\cite{cosm} it leads to strong coupling in perturbations~\cite{cosmpert}.  In our
case, a sensible stable and weakly coupled theory exists as long as $\weff\neq-1$.

\medskip

\paragraph{Impact on the static gravitational potential.} Having the general expressions of the
masses~(\ref{masses}) in the Lorentz breaking (LB) case, we can study the consequences of the
Bianchi identity and the existence of a nontrivial cosmology for the linearized gravitational
potential. The gravitational potential is modified at large distances and amounts to a combination
of two Yukawa terms~\cite{Rubakov,usweak,uslong,Yuk1,Yuk2}:
\be
\Phi=\frac{G\,T_{00}}{2r} \Big(A_1\,{\rm e}^{-M_1 r}+A_2\,{\rm e}^{-M_2 r} \Big)\,,
\ee
with $A_1+A_2=1$, which implies that in the short distance limit $r\ll M_{1,2}$ the potential
reduces to the Newtonian expression (absence of vDVZ discontinuity).  The masses $M_1$, $M_2$ and
the coefficients $A_1$, $A_2$ are given in terms of the LB masses, and their full expressions are
not particularly illuminating.

What is interesting is that, after the discussion above, $\U'\sim \weff+1$ has to be sensibly smaller than 1 if
cosmology requires $\weff$ to be near $-1$.  In this limit, we have (in the $m>H,H'$ hypothesis)
\bea
 M_1^2&\simeq& m^2\,\frac{3\, \U'}{1 + x }+O\big({\U'}^2\big) 
\,,\qquad  A_1\simeq  \frac{ x}
              {\U'}+O(1)\,, \\
 M_2^2&\simeq&m^2\, \frac{3\, \U'}{1 - x}+O\big({\U'}^2\big)
\,,\qquad  A_2\simeq -\frac{ x}
              {\U'}+O(1)\,,
\eea
where $x\equiv\sqrt{1+3 \l_{23}/2\l_2}\sim O(1)$. The conclusion from the above $\U'\ll1$ limit is
that if one wishes to increase $m$ to be larger than the Hubble scale, i.e.\ relevant for
phenomenology, and at the same time keep $\weff\simeq -1$, then the Yukawa distance scale is pushed
again at the Horizon scale.  Thus the requirement of approximately deSitter phase generically
hinders a possible large distance modification of gravity.

The only corner in parameter space which may lead to nontrivial modifications of the gravitational
potential at distances smaller than the Hubble scale can be reached with a tuning for $x\simeq 1$,
in which case $M_2\gg M_1$ and $A_1$, $A_2$ are large and of opposite sign. In this case, discussed
in~\cite{usweak}, the gravitational potential is Newtonian at short distances and Yukawa at very
large distances, but there is a rise at intermediate radii which can mimic the presence of
additional (dark) matter.  The limit $x\simeq 1$ corresponds to $\l_{23}\simeq0$, and notice that
the Schwarzschild spatial cutoff discussed above can become larger, in this limit.

\section{Conclusions}
\label{conc}

In this work we analyzed spatially flat FRW cosmology of  massive gravity theories with five
propagating degrees of freedom.  The analysis is model independent and is based on the powerful
nonperturbative results of~\cite{uscan,usweak,uslong}, which enable one to express the deforming
potential in terms of two functions $\U$ and $\E$. 

A first important result is that the existence of a nontrivial spatially flat FRW cosmological background requires,
due to the Bianchi identities, a stringent condition on the potential, which selects a particular
subclass of theories.  The Lorentz-invariant DeRGT potential~\cite{Gabadadze:2011} is not among
those; as a result Lorentz-breaking in the gravitational sector is a consequence of requiring FRW
cosmology to exist.

In this subclass which admits a nontrivial cosmology, the massive deformation of GR appears first of
all as an effective ``gravitational'' perfect fluid with energy density and pressure determined
solely by $\U$, with an effective equation of state $\weff=-1-\U'/2\U$. For instance, it can mimic
dark energy when $2 \U'/\U <1$. Quite generally thus, massive gravity can easily account for the
present acceleration of the Universe, possibly with a varying equation of state.

The study of static perturbations in these theories confirmed that from a phenomenological point of
view the potentials with Lorentz breaking perform much better than the Lorentz invariant ones.  The
key point is that all the five required modes receive a kinetic term at linear level and are thus
weakly coupled.  In addition the classical strong coupling scale is screened in the weak-field
expansion and one can use perturbation theory around a static source much like in GR: the scale at
which the weak field expansion breaks down is the same as in GR, the Schwarzschild radius, and there
is no Vainshtein strong coupling scale or phenomenon.  If quantum effect are taken into account, an
energy scale $\sim\Lambda_2=(mM_{Pl})^{1/2}$ is the expected quantum cutoff and again the
Lorentz-breaking theories are much better off with respect to the Lorentz invariant ones (where the
cutoff is $\Lambda_3=(m^2M_{Pl})^{1/3}$). In case the graviton mass is taken at cosmological scales,
$\Lambda_2\simeq(10^{-4}\,\text{mm})^{-1}$ (while $\Lambda_3$ is of the order of $1000$\,Km).  Thus
in these theories quantum effects are automatically confined at submillimiter scales, possibly
tested by future short-distance gravity probes.

The study of perturbations also allowed us to discover a general link between the cosmological
background and their behaviour. We have shown that, if strict Minkowski space is a vacuum solution,
then gravitational perturbations are strongly coupled, because the temporal kinetic terms of the
vector and scalar perturbations vanish in this background. The same happens in strict de Sitter
space. Therefore, some deviation from maximally symmetric backgrounds is required to have an healthy
and calculable theory.  It is remarkable that, in a quasi-dS universe dominated by the induced
gravitational dark energy, one can find a simple relation between the cutoff scale of the theory and
the deviation of $\weff$ from $-1$, as $\Lambda\simeq 4\pi \sqrt{HM_{Pl}}(\weff+1)^{1/4}$.  Thus, the
requirement of the of absence strong coupling in the present quasi-dS phase can be used to predict
the equation of state of dark energy.

\smallskip

Finally, and more broadly, it is a natural question whether, once flat-background Lorentz invariance
is not imposed in the gravitational sector, there exist other massive deformations of gravity with a
number of DoF different from five. The answer is positive and we will report on their general
features in a separate publication~\cite{usnext}.

\begin{appendix}

\section{Quadratic Action}
\label{appquad}
The starting point is the following perturbed FRW background
\be
\begin{split}
& g_{\mu \nu} = \bar g_{\mu \nu} +a^2 \,  h_{\mu \nu} \, ,  \qquad
h_{00} = a^2 \left( N^i N^j \bar \g_{ij}-2 \bar N \, \delta N - \delta
  N^2  \right) \, , \\
&h_{0i}=
a^2 \left( \bar \gamma_{ij} + \a_{ij} \right) \, N^j \, , \qquad
h_{ij} =a^2 \, 
\a_{ij} \, .
\end{split}
\label{rel}
\ee
The  of perturbations  can be decomposed as
\be
\begin{split}
&h_{00} = \psi \, , \qquad h_{ij}= \chi_{ij} + \de_i s_j + \de_j s_i + \delta_{ij} \, \tau +
\de_i \de_j \sigma , \qquad h_{0i} = u_i +
\de_i v  \, ; \\
&  \chi_{ij} \delta^{ij} = \de_j  \chi_{ij} = \de_i s_i = \de_i u_i =0
\, .
\end{split}
\ee
We have chosen $ N=a$. The total action is given by
\be
S =  \plm \int d^4x \, \sqrt{g}  ( R- m^2 \, V) \equiv   \plm \, \int
d^4x \,  \sqrt{g}\left[ R-6\, H^2-(m^2 \, V -6 \, H^2) \right] \, .
\ee
Using (\ref{rel}), the quadratic expansion of $V$, shifted by $6 \, H^2$, can be
written as 

\be
m^2 \sqrt{g} (V-6 \, H^2) =a^4 \, \left[ t^{\mu \nu}_V 
h_{\mu \nu} + \frac{1}{4} \left(m_0^2 \, h_{00}^2+2\,m_1^2\, h_{0i}
    h_{0i}   -2\,m_4^2 \,h_{00}\,  h_{ii} +m_3^2 
\,  h_{ii}^2  - m_2^2\, h_{ij}\; h_{ij}\right) \right] \, .
\ee
The values of the various masses can be computed for any $V$ of the
form (\ref{pot}), the result is the following
\be
\begin{split}
& m_0^2 = 0\, , \qquad m_1^2= m^2
\, 2 \, a \, \bar \U'  \, \X 
\, , \qquad
m_2^2 =2 m^2 \left(\frac{\bar \E''_t}{a} + \bar \U''_t + 2 \, \bar \U' - \frac{ \bar
    \E}{2 a}  \right) \, , \\
&m_3^2 = 2 m^2 \left( 
  \, \bar \U' + \frac{\bar
    \E}{4 \, a} -\bar \U''_s -\,  \frac{\bar \E''_s}{a}  \right)\, ,
\qquad m_4^2 = m^2 \, \bar \U' \, ; \\
&  t^{00}_V =  \frac{m^2}{2} \ \, \bar \U \, , \qquad 
t^{ij}_V = m^2 \left(\bar \U'-\frac{ \bar \U}{2} \right) \delta^{ij} \, ;
\end{split} 
\label{lambdas}
\ee
where
\be
\begin{split}
&\frac{\de \U}{\de \K^{ij}}_{| \bar g}= \bar \U' \, \bar \g_{ij} \, ,
\qquad \frac{\de^2 \E}{\de \xi^i \de \xi^j}_{| \bar g}=\bar \E'' \,
\bar  \g_{ij} \, , \qquad  \frac{\de \xi^m}{\de N^i}_{| \bar g}
\frac{\de \xi^n}{\de N^j}_{| \bar g} \bar \g_{mn} = \bar \g_{ij} \,
\kappa \, , \\
&  \frac{\de^2 \E}{\de \g_{ij} \de \g_{mn}}_{| \bar g}= \bar \E''_s
\, \g^{ij}  \g^{mn}+\bar \E''_t \,   \ha \, \left(\g^{im} \g^{jn} +\g^{in} \g^{jm}
\right) \, ;\\
& \frac{\de \U}{\de \K^{ij} \de \K^{mn}}_{| \bar
  g}= \U''_s \,  \g_{ij}  \g_{mn}+ \bar \U''_t \,  \ha \left(\g_{im} \g_{jn} +\g_{in} \g_{jm}
\right) \, ;\\
\, 
&\ha \left( \bar \g_{ik} \frac{\de \xi^k}{\de N^j}_{| \bar g}+
\bar \g_{jk} \frac{\de \xi^k}{\de N^i}_{| \bar g} \right) = \bar
\g_{ij} \, \X\, .
\end{split}
\ee
Background quantities are denoted by a bar. In (\ref{lambdas}) the background Bianchi identity (\ref{bianchi}) has been used to
eliminate $\bar \E'$. 
The value of $\X$ can explicitly determined 
expanding $\xi^i$ in powers of $N^j$, namely 
\be
\xi^i = A^i_j \, N^j + B^i_{jm} \, N^j N^m + C^i_j \, N^j \, \delta N \cdots \, ;
\ee
thus 
\be
\frac{\de \xi^i}{\de N^j}_{|\bar g} = A^i_j \, . 
\ee
From $Q^j \U_{ij} =- \E_i$ and using that $\U_{ij} = -2
\bar \U' \bar \gamma_{ij} + O(\xi)^2$ we have  that
\be
\frac{\de \xi^i}{\de N^j}_{|\bar g} =\delta^i_j \left(a
  +\frac{\bar \E''_v}{2 \bar \U'} \right)^{-1} \,  ;
\ee
thus 
\be
\X = \left(a
  +\frac{\bar \E''_v}{2 \bar \U'} \right)^{-1} \, .
\ee
In particular 
\be
\X_{|Q=0} = a^{-1} \, .
\ee
The quadratic expansion of total action for an Universe dominated
by dark energy induced by massive gravity modification can be written as
\be
\begin{split}
&
S_{EH} = \int d^4x \, \sqrt{g} \, \plm \, R =  \int d^4x \, \plm
\left[L_{(0)} +L_{(1)} + L_{(2)} + \cdots  \right ]\; ; \\
& L_{(1)}= L_{(1)}^{(s)} + L_{(1)}^{(v)}+  L_{(1)}^{(t)} \, , \qquad
L_{(2)}= L_{(2)}^{(s)} + L_{(2)}^{(v)}+  
L_{(2)}^{(t)} \, .
\end{split}
\label{quad}
\ee
The tensor part  is given by
\be
\begin{split}
&L_{(2) \, \text{tot}}^{(t)}= \frac{a^2}{2} \,  \chi_{ij}'
\chi_{ij}' + \frac{a^2}{2} \, 
 \chi_{ij} \left( \Delta + a^2 \,m_2^2 
   \right) \chi_{ij}\, ; \\
\end{split}
\ee  
For the vectors we get
\be
\begin{split}
&L_{(2) \, \text{tot}}^{(v)}= - \frac{a^2}{2} (u_i -s_i')\Delta  (u_i -s_i') + 3 \,
\frac{a^4}{2} \, m_1^2 \,  u_i u_i- \frac{a^2}{2} \, s_i \Delta s_i \,
m_2^2  \, ; \\
\end{split}
\ee
Finally, for scalars we have 
\be
\begin{split}
&L_{(1) \, tot}^{(s)}=  a^2 \, \left( \frac{m^2 \, a^2 \, \bar \U}{2}- 3 \H^2
  \right) \, \psi + a^2 \left[ a^2 m^2(\bar \U' -\frac{\bar
      \U}{2})+\H^2 + 2 \H' \right] 3 \, \tau \, ;\\
&L_{(2) \, tot}^{(s)}=\frac{a^2}{4}\Big\{-6(\tau'+\H \psi)^2+
2(2\psi-\tau)\Delta \tau+4(\tau'+\H \psi)\Delta (2v- \s')\\
&\qquad\qquad\qquad+a^2\Big[m_0^2 \psi^2-2 m_1^2v\Delta v-m_2^2(\s \Delta^2\s+2\tau\Delta\s+3\tau^2)\\
&\qquad\qquad\qquad\qquad+m_3^2(\Delta\s+3\tau)^2-2m_4^2\psi(\Delta\s+3\tau)\Big]\Big\}\, .
\end{split}
\ee
We have set $8 \pi G= 1/(2 \plm)$.  Notice that the linear term in the scalar action gives the
background equations of motion. The tensor, vector and scalar parts of the action precisely
coincides with ones studied in \cite{diego} and basically we can use all results in there using the
values (\ref{lambdas}) for the masses (the only difference in comparison with~\cite{diego} is that
$m_2^2$ and $m_3^2$ are shifted in that work).

\end{appendix}

\end{document}